    \documentclass[preprint,12pt]{elsarticle}
    \usepackage{amssymb}
    \usepackage{amsmath}
    \usepackage{tcolorbox}
    \usepackage[utf8]{inputenc}
    \usepackage{subfig}
    \usepackage{booktabs} 
    \usepackage{url}
    \usepackage{array}
    \usepackage{tabularx}
    \usepackage{caption}
    \usepackage{algorithm}
    \usepackage{algorithmic}
    \usepackage{listings}
    \usepackage{xcolor}
    
    \usepackage{hyperref}
    \usepackage{soul}
    \usepackage{tcolorbox} %
    \journal{Information and Software Technology}
    
\begin{document}
    
    \begin{frontmatter}
    
    %% Title, authors and addresses
    
    %% use the tnoteref command within \title for footnotes;
    %% use the tnotetext command for theassociated footnote;
    %% use the fnref command within \author or \affiliation for footnotes;
    %% use the fntext command for theassociated footnote;
    %% use the corref command within \author for corresponding author footnotes;
    %% use the cortext command for theassociated footnote;
    %% use the ead command for the email address,
    %% and the form \ead[url] for the home page:
    %% \title{Title\tnoteref{label1}}
    %% \tnotetext[label1]{}
    %% \author{Name\corref{cor1}\fnref{label2}}
    %% \ead{email address}
    %% \ead[url]{home page}
    %% \fntext[label2]{}
    %% \cortext[cor1]{}
    %% \affiliation{organization={},
    %%             addressline={},
    %%             city={},
    %%             postcode={},
    %%             state={},
    %%             country={}}
    %% \fntext[label3]{}
    
    \title{ A Context-Driven Approach for Co-Auditing Smart Contracts Using GPT-4 Code Interpreter}
    
    %% use optional labels to link authors explicitly to addresses:
    %% \author[label1,label2]{}
    %% \affiliation[label1]{organization={},
    %%             addressline={},
    %%             city={},
    %%             postcode={},
    %%             state={},
    %%             country={}}
    %%
    %% \affiliation[label2]{organization={},
    %%             addressline={},
    %%             city={},
    %%             postcode={},
    %%             state={},
    %%             country={}}
    
    \author[poly]{Mohamed Salah Bouafif}
    \author[polyhong]{Chen	Zheng}
    \author[icelandUniv]{Ilham	Qasse}
    \author[quant]{Ed	Zulkoski}
    \author[poly,icelandUniv]{Mohammad	Hamdaqa}
    \author[poly]{Foutse Khomh}
    %% Author affiliation
    \affiliation[poly] { 
      organization={Department of Computer and Software Engineering, Polytechnique Montreal},
      city={Montreal},
      country={Canada}
    }
    
    \affiliation[icelandUniv] {
      organization={Department of Computer Science, Reykjavik University},
      city={Reykjavík},
      country={Iceland}
    }
    
    \affiliation[polyhong] {
      organization={Department of Computer Science, The Hong Kong Polytechnic University },
      city={Hung Hom},
      country={Hong Kong}
    }
    
    \affiliation[quant] {
      organization={Quantstamp},
      city={Dover},
      country={United States}
    }
    
    %% Abstract
    \begin{abstract}
    %% Text of abstract
    \noindent\textbf{Context}: The surge in the adoption of smart contracts necessitates rigorous auditing to ensure their security  and reliability. Manual auditing, although comprehensive, is time-consuming and heavily reliant on the auditor's expertise. With the rise of Large Language Models (LLMs), there is growing interest in leveraging them to assist auditors in the auditing process (co-auditing). However, the effectiveness of LLMs in smart contract co-auditing is contingent upon the design of the input prompts, especially in terms of context description and code length.
    
    \noindent \textbf{Objectives}: This paper introduces a novel context-driven prompting technique for smart contract co-auditing. The primary objective is to improve the effectiveness of LLMs in detecting vulnerabilities in smart contracts through enhanced prompt design.
    
    \noindent \textbf{Methods}:
    Our approach employs three techniques for context scoping and augmentation, namely: \textit{Code Scoping}, which involves chunking long code into self-contained code segments based on code interdependencies; \textit{Assessment Scoping}, which enhances context description based on the target assessment goal to limit the search space; and \textit{Reporting Scoping}, which forces a specific format for the generated response.
    
    \noindent\textbf{Results}: Through empirical evaluations on publicly available vulnerable contracts, our method demonstrated a detection rate of 96\% for vulnerable functions, outperforming the native prompting approach, which detected only 53\%. To assess the reliability of our prompting approach, manual analysis of the results was conducted by two expert auditors from Quantstamp, a world-leading smart contract auditing company. The experts' analysis indicates that, in unlabeled datasets, our proposed approach enhances the proficiency of the GPT-4 code interpreter in detecting vulnerabilities.
    
    \noindent \textbf{Conclusion}: The context-driven prompting technique significantly improves the effectiveness of LLMs in smart contract co-auditing. This method not only enhances the detection rate of vulnerabilities but also reduces the dependency on manual auditing, thereby making the auditing process more efficient and reliable.
    \end{abstract}

    %% Keywords
    \begin{keyword}
    %% keywords here, in the form: keyword \sep keyword
    Empirical Study \sep Smart Contracts\sep Auditing\sep Large Language Models \sep ChatGPT \sep Vulnerabilities.
    %% PACS codes here, in the form: \PACS code \sep code
    
    %% MSC codes here, in the form: \MSC code \sep code
    %% or \MSC[2008] code \sep code (2000 is the default)
    
    \end{keyword}
    
    \end{frontmatter}
    
    %% Add \usepackage{lineno} before \begin{document} and uncomment 
    %% following line to enable line numbers
    %% \linenumbers
    
    %% main text
    %%

    \section{Introduction}

    Smart contracts have become the foundation of automation and trust through secure and transparent transactions \cite{de2019old}. They are adopted in various industries, from finance to real estate to supply chain management, promising efficiency, transparency, and trust in a trustless environment~\cite{zheng2020overview, kolvart2016smart}.
    However, the growth in adoption of smart contracts brings along challenges. Smart contracts are vulnerable to exploits, misuse, and unforeseen vulnerabilities due to the very features that make them powerful (such as immutability and transparency) \cite{zhang2020framework}. This underscores the necessity to audit smart contracts, a process that examines the code and features of the contracts to identify and mitigate potential risks \cite{he2020smart}.
    
    Smart contract auditing is typically performed manually by specialized auditing engineers and requires a deep understanding of the business use case and the historical attack events \cite{zou2019smart}. Auditing smart contracts is time-consuming and challenging, contingent upon the auditor's expertise and experience \cite{zou2019smart}. Auditors require a diverse skill set that encompasses proficiency in debugging, a deep understanding of various code analysis techniques, knowledge of blockchain platforms (e.g., EVM), and a grasp of the business domain and rules embedded within the protocol's code \cite{zou2019smart}.
    
    A possible solution to address these challenges is using Large Language Models (LLMs) for smart contract auditing.
    LLMs, empowered by their vast pre-trained knowledge, have demonstrated remarkable capabilities in understanding and generating code \cite{zheng2023towards}. Recently, different studies focused on exploring the usage of LLMs for automating code review, including smart contracts \cite{sarsa2022automatic}. Nevertheless, LLMs can be susceptible to the phrasing and length of the context in the input prompts. Subtle phrasing, tone, or context changes of a prompt can lead to significantly different model output, especially when the project's source code is very long and the dependencies are very complex. As shown in a recent study by David et al. \cite{david2023you}, increasing the length of the prompt context reduced the performance of the LLM in detecting vulnerabilities. Similarly, Chen et al. \cite{chen2023chatgpt} suggest that GPT models may encounter detection failures in long smart contract code due to inherent token limitations. Therefore, an appropriate prompt design approach that takes context phrasing and length into account is necessary in order to assess the LLMs for detecting vulnerabilities and generating appropriate audit reports. 
    
    In this work, we introduced a novel context-driven prompting technique for smart contract auditing that aims to provide appropriate wording and enough code context based on direct code dependencies to uncover vulnerabilities and provide code audit recommendations. To guide the LLM in identifying the vulnerabilities, we used the Common Weakness Enumeration (CWE)\cite{david2023you} to reduce the search space, in addition to a set of predefined Common Audit Questions (CAQ) extracted from previous audit reports to scope the generated reports. Moreover, to provide sufficient code context while avoiding including the entire code within the prompt, we developed a chunking approach based on code call graphs, which we refer to as Code Call List or CCL-based chunking. For evaluation, we conduct experiments to answer the following research questions:  
    
    \begin{enumerate}
    \item \textbf{RQ1. How does CCL-based chunking perform in comparison to full code prompting?}
    The concept of chunking, particularly CCL-based chunking, is inspired by the manual auditing process employed by expert auditors. During the audit of a smart contract, auditors meticulously verify each function to ensure it is free from vulnerabilities. This necessitates a comprehensive understanding of all the code executed within each code snippet. For Research Question 1 (RQ1), we assess the performance of CCL-based chunking relative to full-code prompting. In full-code prompting, the GPT-4 code interpreter typically executes automated mechanical chunking by dividing the code based on the maximum token size permissible for a single prompt. Our evaluation aims to determine the effectiveness of CCL-based chunking in comparison to this automated approach. 
    \item \textbf{RQ2. What is the impact of different prompt phrasing approaches (CWE-based, or CAQ-based) on the effectiveness of the GPT results of CCL-based chunking?}
    The objective of RQ2 is to study the impact of utilizing different wording and phrasing techniques for context augmentation based on expected results (e.g., the expected vulnerability in CWE-based) or generic auditing inquiries (searching for all vulnerabilities in CAQ-based) approaches on the cost and effectiveness of CCL-based chunking to guide the ChatGPT to retrieve the correct results. 
    \item \textbf{RQ3: To what extent does a human auditor's analysis of GPT-4 code interpreter's results on an unlabeled dataset align with our previous findings?}
    The objective of RQ3 is to evaluate the reliability of our proposed prompting approach by assessing its consistency with manual auditing by expert auditors from our partner (a world-leading smart contract auditing company). A second experiment was conducted using unlabeled datasets, and expert auditors were invited to perform an independent analysis and qualitative assessment of GPT-4 code interpreter's responses, providing an additional layer of verification to our proposed methodology.
    \end{enumerate}

    The main contribution of the paper is as follows:

    \begin{enumerate}
        \item We introduce a context-driven approach to enhance LLM performance in smart contract auditing by proposing three scoping techniques: Code Scoping, Assessment Scoping, and Reporting Scoping. These techniques aim to improve understanding and generate accurate audit reports by addressing the challenges of (i) long code contexts in prompts, (ii) context related to the audit process, and (iii) context related to expected target responses.
        \item For each scoping technique, we provide a practical implementation. Code Scoping is realized through CCL (Code Call List), which organizes code into logical chunks based on function call dependencies to improve contextual awareness. Assessment Scoping is realized through CAQ (Common Audit Questions), a tailored set of questions that guide the LLM's focus on specific vulnerabilities. Reporting Scoping is realized through CWE (Common Weakness Enumeration), helping the LLM categorize and report vulnerabilities using standardized industry classifications.
        \item We present hypotheses regarding the effectiveness of the scoping techniques and empirically test these through experiments to assess their impact on audit quality.
        \item We provide a new dataset of vulnerable functions, manually labeled by two professional smart contract auditors, to aid in the evaluation and validation of the proposed methodologies.
    \end{enumerate}

    \begin{comment}

    \begin{enumerate}
        \item We introduce a novel context-driven prompting technique for smart contract co-auditing that aims to provide appropriate wording and enough code context based on direct code dependencies to provide code audit recommendations.
        \item We present a dataset of vulnerable functions and their corresponding Code Call Lists, annotated by experienced auditors, facilitating empirical evaluation and future research.
        \item We empirically evaluated the different prompt engineering techniques for context representation and we asses the correctness and relevance of the generated code audits.
        \item We provide the first labeled dataset containing solidity code that have been audited professional auditors.
    \end{enumerate}
     \end{comment}
    
    The rest of this research paper is organized as follows: Section II provides the motivation for this work. Section III explains our methodology. Section IV presents our empirical evaluation. Section V discusses our findings. Section VI provides all the artifacts used for this work and section VII introduces related work. Section VIII outlines the threats to validity and Section IX concludes the paper.

    \section{Smart Contract Co-Auditing}
    \label{sec:motivation}
    \subsection{Code Auditing}
    Software development encompasses a range of software quality assurance activities, including defect management, testing, and code review \cite{fan2023large, hou2023large}. In the context of code review, reviewers evaluate whether the source code meets both functional and non-functional requirements. On the other hand, blockchain-based projects also require a rigorous quality assurance process to ensure the quality, security, and overall reliability of the codebase. This process is called code audit. Code auditing and code review are both critical processes in software development. While they share similarities, they differ in scope, purpose, and complexity \cite{he2020smart}.
    In the context of smart contracts, code auditing can be a complex process that involves a comprehensive and in-depth analysis of the whole codebase of the project with the objective of identifying vulnerabilities, potential attack vectors, and security weaknesses. This process involves different tasks, including fault localization, issue explanation, documentation, and issue fix suggestions, which can be time-consuming and highly dependent on the qualifications of the auditors.

    \subsection{Co-auditing with Generative AI}
    Co-auditing refers to the practice of involving multiple auditors, which can include both artificial intelligence (AI) systems and human oversight, in the auditing process. \cite{zheng2023towards}. A practice that witnessed fast advancement with the emergence of LLMs. The need for co-auditing can vary depending on the domain, but it is crucial in applications where factual errors can have serious repercussions \cite{joshi2023repair}. An example of such an area is smart contract auditing.
    Recently, different studies have focused on exploring the use of LLMs for code auditing \cite{hou2023large, chen2023chatgpt}. While these works highlight the capacity of such models for Solidity code understanding and interpretation, the accuracy of LLMs for the task of detecting vulnerabilities is still low. In \cite{chen2023chatgpt}, different limitations have been reported for the use of GPT-4 code interpreter for code audits, including uncertainty and context length. The performance of GPT-4 code interpreter can degrade when the prompted code becomes very long. Long contextual dependencies can lead to ambiguity in properly understanding the prompted codebase.
    Besides, code auditing requires knowledge about the whole code base, which makes it more complex for tools such as GPT. 
    \begin{comment}
        
    Although, theoretically, the standard GPT-4 code interpreter model offers 8000 tokens as a context limit, the paper argues that the maximum code size supported for the task of vulnerability detection is much lower, and the performance of GPT-4 code interpreter degrades considerably when the lines of code exceed a certain limit that is lower than the maximum token limit supported by GPT-4 code interpreter. 
    \end{comment}
    
    Despite the theoretical capacity of large language models (LLMs) like GPT-4 to process up to 8000 tokens, our research reveals that their efficacy in smart contract vulnerability detection is hindered by limitations in contextual understanding, particularly when encountering complex code structures. Notably, the performance degradation of LLMs in this domain is not solely attributable to context length, but rather a multifaceted issue encompassing the model's ability to capture nuanced code relationships, reason about indirect dependencies, and generalize to novel contract architectures. 
    The manual tests we performed revealed that GPT-4 code interpreter fails to detect all the vulnerabilities in a smart contract when the codebase is too long, inducing many false negatives. This underscores the need for innovative prompting techniques that can effectively harness the strengths of LLMs while mitigating their limitations in smart contract auditing.
    The integration of generative AI in code auditing poses a unique challenge, as highlighted by Liu et al. in "What It Wants Me To Say" \cite{liu2023wants}. The disconnect between the end-user programmer's requirements and the output of large language models (LLMs) can lead to suboptimal results in tasks like code generation and vulnerability detection. In the context of code auditing, this gap can have significant implications, as LLMs may not fully comprehend the nuanced security concerns and auditing standards that human auditors take for granted. However, LLMs can still play a vital role in augmenting the code auditing process, particularly in tasks such as:
    \begin{itemize}
    
    \item Identifying potential vulnerabilities and anomalies in large codebases
    \item Providing recommendations for code improvements and security enhancements

    \end{itemize}

    The authors emphasized the role of scoping, elaboration, language restructuring, and intent shaping in grounded utterance \cite{liu2023wants}. Stemming from these results, this work focuses on approaches to improving co-auditing through context scoping and augmentation.  We introduce the following three scopes: 
    
    \begin{enumerate}
    \item \textbf{Code scoping}: This scoping is related to managing long code with complex inter-dependencies. In such cases, scoping should include methods to direct the LLM's attention to the interrelated code elements that could be far apart but have close interactions. Understanding how code elements interact could lead to identifying potential vulnerabilities. While we do not explicitly quantify this relationship in the paper, the concept involves recognizing and structuring the context based on inter-dependencies, which can be informed by function call graphs and dependency analysis as explained in Section 3.1.
    \item \textbf{Reporting scoping}: This scoping is related to the output audit report; an auditor needs to report specific content in a specific order and format. This is highly dependent on the auditing process. Scoping this information through specific questions not only guides the LLM to generate the required information but also focuses attention on the next tokens that need to be generated, which could positively influence the output results.
    \item \textbf{Assessment scoping}: This scoping is related to defining the specific vulnerabilities, weaknesses, or security issues an auditor is targeting within the codebase. For example, the auditor may target issues like reentrancy, integer overflows, access control vulnerabilities, or gas limit issues. This focus helps narrow down the scope of the context.  This scoping technique is closely related to retrieval-augmented approaches, as it helps guide the LLM to focus on the most relevant information by retrieving context-specific details for accurate vulnerability detection. While we do not explicitly define formal parameters or metrics for Assessment Scoping at this stage, in the implementation of this scoping technique in section xx the context was identified based on targeted vulnerabilities retrieved from the Common Weakness Enumeration (CWE) database and evaluated by accuracy of vulnerability detection 
    
    \end{enumerate}

    Researchers have been investigating various aspects of context in large language models (LLMs), including how it is defined, utilized, and evaluated \cite{li2023loogle, krishnamurthy2024can}. In \cite{li2023loogle}, authors address the challenge of long-context understanding for LLMs.  The paper defines the context window as the maximum amount of text or tokens that a model can process in a single input. The study revealed that LLMs excel in short dependency tasks but struggle with more complex long dependency tasks, indicating a gap in true long-context understanding. Yilun Zhu et al. \cite{zhu2024can} conducted an empirical investigation into the contextual comprehension capabilities of large language models (LLMs). Their systematic evaluation assessed the ability of LLMs to accurately interpret subtle and implicit contextual information, a crucial factor in tasks requiring profound contextual awareness. The findings revealed that while LLMs exhibit satisfactory performance in general linguistic tasks, they frequently misinterpret or fail to comprehend nuanced contextual cues.
    This study underscores the significance of contextual definition and structure in optimizing LLM performance, particularly for code generation and vulnerability detection tasks. Notably, LLM performance degrades when processing large codebases, characterized by intricate dependencies and numerous function calls, even when token counts remain within contextual windows \cite{levy2024same}.
    For comprehensive codebase understanding, semantic relationships between disparate code segments assume paramount importance in structuring input prompts. Harnessing the structural properties of function call graphs, throw code scoping, offers a promising approach to represent logical connections between functions, thereby enhancing contextual awareness.
    
    To inform the development of effective Large Language Model (LLM) auditing queries for smart contracts, we employed a mixed-methods approach combining card sorting methodology and expert elicitation. Our investigation involved analyzing a dataset of audit reports extracted from Code4Arena, a prominent capture-the-flag platform utilized by the auditing community, to identify key themes, patterns, and vulnerabilities prevalent in smart contract audits. In collaboration with professional auditors, we distilled these findings into a comprehensive set of tailored questions designed to elicit relevant information from the LLM, ensuring alignment with industry auditing standards. By integrating human expertise and real-world audit report data, our approach establishes a robust foundation for exploring the potential of LLMs in automated smart contract auditing, providing a systematic and informed basis for enhancing the efficiency and effectiveness of this critical process.
    
    This study integrates Large Language Models (LLMs) into real-world smart contract auditing workflows, enhancing auditing practice by tailoring LLM outputs to specific vulnerability detection and reporting requirements. By bridging the gap between LLM general capabilities and smart contract security needs, our approach addresses specifically auditing smart contracts. Our focused methodology provides actionable insights into LLM applicability, rather than broadly investigating LLM behavior.
    
    In the evaluation section, we provide an example of implementation of our understanding of the context definition for the task of auditing smart contract.

    \subsection{Motivation Example}
    
    \label{ME}
    Listing \ref{lst1} illustrates an example of a smart contract called TokenizingVault. This vault contract takes users' diverse ERC20s (token standard) deposits and then issues corresponding "credits" as deposit tokens. These tokens can then be exchanged to reclaim the initial deposit. This design provides fungibility that allows users to trade their secured assets seamlessly.
    The TokenizingVault contract imports two other contracts, namely ERC20.sol and ERC20CreditToken.sol. ERC20 and ERC20CreditToken also import different other contracts which we tried to prompt this contract along with the other imported contract. The contracts source code length exceeds 8000 token supported by GPT4. A direct solution for such a challenge is  a naive truncation. We prompted the source code of the contracts sequentially in different prompts. We used a binary classification of the contract and asked GPT to locate the vulnerable functions, if any, and explain the vulnerability.
    \begin{lstlisting}[caption= TokenizingVault contract,
      label=lst1]
    pragma solidity 0.8.15;
    
    import "@openzeppelin/contracts/token/ERC20/ERC20.sol";
    import {ERC20CreditToken} from "./ERC20CreditToken.sol";
    
    contract TokenizingVault {
        /// ERC20 accounting tokens (unique per underlying)
        mapping(ERC20 => ERC20CreditToken) public creditTokens;
        /// ERC20CreditToken reference implementation
        ERC20CreditToken public immutable creditTokenImpl;
        constructor() {
            creditTokenImpl = new ERC20CreditToken();
        }
        function create(ERC20 underlying, uint256 amount)
            external nonReentrant returns (ERC20CreditToken, uint256)
        {
            ERC20CreditToken creditToken = creditTokens[underlying];
    
            // Revert if no token exists, must call deploy first
            if (creditToken == ERC20CreditToken(address(0x00)))
                revert('Token Does Not Exist');
    
            // Transfer in underlying
            underlying.transferFrom(msg.sender, address(this), amount);
    
            // Mint new credit tokens
            creditToken.mint(msg.sender, amount);
    
            return (creditToken, amount);
        }
        function redeem(ERC20CreditToken token, uint256 amount)
            external nonReentrant
        {
            token.burn(msg.sender, amount);
            token.underlying().transfer(msg.sender, amount);
        }
        function deploy(ERC20 underlying)
            external nonReentrant returns (ERC20CreditToken)
        {
            // Create credit token if one doesn't already exist
            ERC20CreditToken creditToken = creditTokens[underlying];
            if (creditToken == ERC20CreditToken(address(0))) {
                bytes memory tokenData = abi.encodePacked(underlying, address(this));
                creditToken = ERC20CreditToken(address(creditTokenImpl).clone(tokenData));
                creditTokens[underlying] = creditToken;
            }
            return creditToken;
        }
    }
    \end{lstlisting}
     Due to the complex dependency, the code base is prompted using a sequence of 5 prompts. GPT is composed of 3 parts: general observations, Potential Vulnerabilities and Suggestions, and conclusions. As we can see in \href{https://chat.openai.com/share/fef0be4f-0def-449a-bf22-30fd9cc373cb}{the generated response}\footnote{https://chat.openai.com/share/fef0be4f-0def-449a-bf22-30fd9cc373cb}, GPT fails to detect a crucial vulnerability which is related to the redeem function.
    
    The core issue for this contract is the function \textit{redeem}. The function \textit{redeem} is external and does not have an access control list or input validation schema. This function is vulnerable because it does not validate if the passed token is authenticated. One exploitation scenario is that an attacker could invoke the redeem function using a maliciously crafted ERC20CreditToken. If this token's \textit{underlying} method points to a legitimate and valuable ERC20 token that other users have stored in the vault, it may lead to unauthorized withdrawals. The vulnerable function is only three lines long. Although GPT's response mentioned different code issues related to the contract, the vulnerability has not been listed. We repeated the experiment various times.
    The potential reasons can be: 
    
    \begin{itemize}
    %\item{\textbf{Inability to understand the business use-case}}:
    %one major challenge for auditing smart contracts is understanding the business use case and the desired function. While prompting the whole source code of the contract can provide the LLMs with a better understanding of the project, it is required that further context and project description should be provided.
    \item{\textbf{Lack of context depth}}:
    In the absence of specific context description, llm is capable of generating outputs in formats that are not uniformly structured.  This irregularity in the generated response presents substantial obstacles in extraction of valuable insights that are both accurate and coherent. 
    The absence of the precise description of the desired output in term of the context inquiry process, related to the questions that needs to be answered, and in term of the generated response, related to expected results template, can lead GPT to generate high level results that can increase the rate of false positive/negative and/or missing certain complex vulnerabilities.
    \item{\textbf{Sensitivity to source code length}}:
    The length of the prompt context and the code can affect the performance of LLMs for vulnerability detection \cite{david2023you, chen2023chatgpt}, where the effective maximum token number for GPT-4 is lower than 8000. 
    \end{itemize}
    This suggests that a more effective chunking technique for the source code is required to fully exploit the LLMs code understanding capabilities. Hence, this paper proposes a new chunking technique based on a function call graph for source code. 
    Moreover, the paper will present comparative experiments involving different groups of prompt engineering techniques for context representation and assess the generated audits, demonstrating the current advantages and limitations of the context length on the performance of LLMs for smart contract audits.

    \section{Implementation of Context-Driven Auditing Method}
    \label{sec:methodology}
\begin{figure}[h!]
    \centering
    \includegraphics[width=0.7\columnwidth]{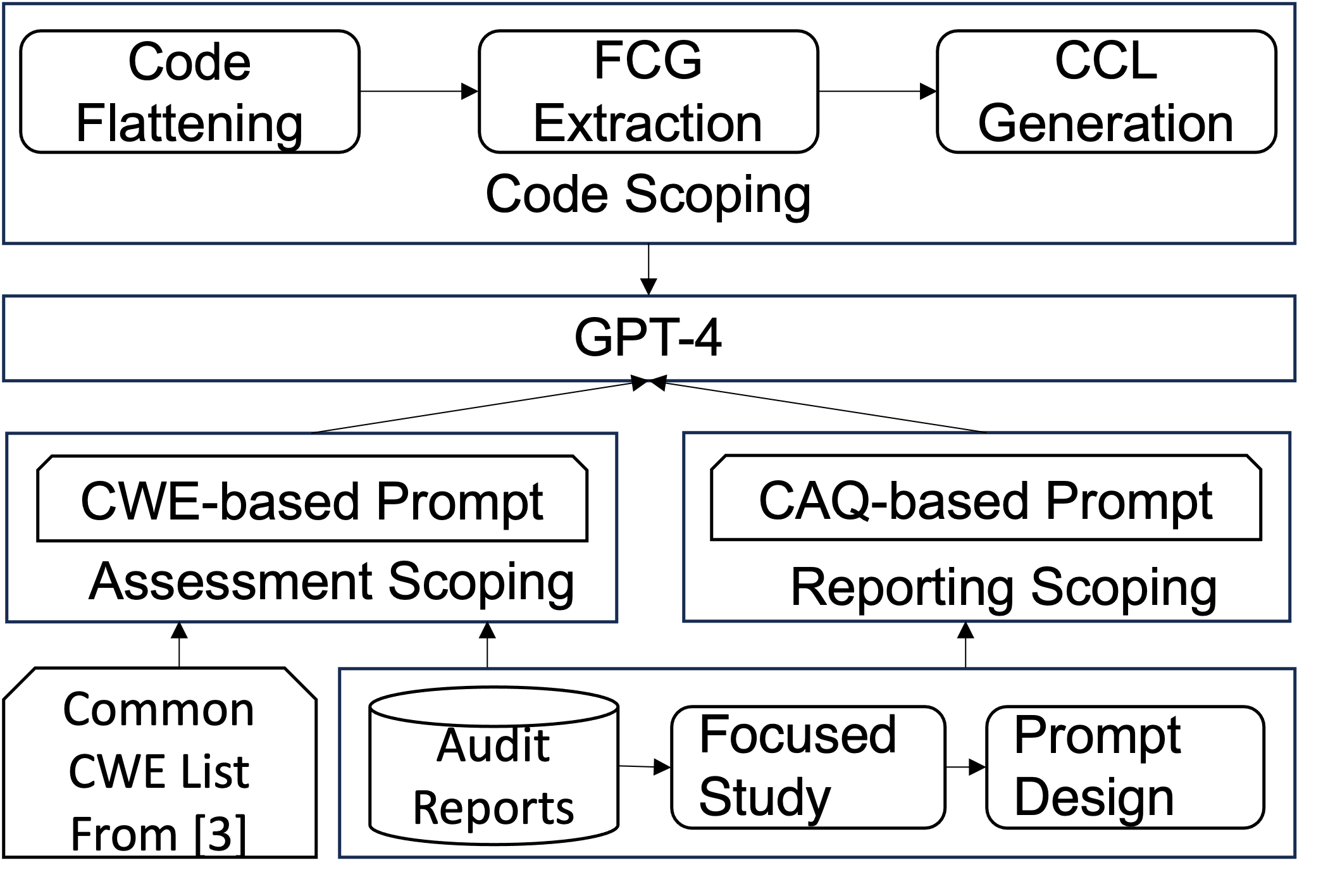}
    \caption{Overview of the study method}
    \label{architecture}
    \end{figure}

    In this paper, we propose three context-augmenting approaches with the goal of addressing the problems of (i) long code contexts in a prompt, (ii) context related to the process of inquiry (i.e., to augment knowledge related to the audit process extracted from previous audit reports), and (iii) context related to the expected target responses (i.e., inquiries about a specific issue or weakness based on the knowledge from previous vulnerabilities). These augmentation approaches can be used to design prompts that can support smart contract co-auditing with the support of GPT. Figure \ref{architecture} illustrates a high-level overview of our approach. The rest of this section explains the three context-augmenting approaches and their corresponding prompt templates. 
    
    \subsection{Code Scoping}
    
    To provide sufficient code context while avoiding including the entire code within the prompt, we propose a chunking approach based on Code Call List (CCL). This process includes code flattening, Function Call Graph (FCG) extraction, and finally, CCL generation.
    
    \subsubsection{Code Flattening}
    
    Smart contract code often involves importing, calling, and referencing other contracts and libraries and implementing interfaces that extend the contract's functionality. These dependencies can increase the complexity of auditing smart contracts, especially when dealing with multiple imported contracts. To simplify this complexity, we employ a code-flattening approach, which consolidates all the code from the contract and its dependencies into a single file. 
    The flattening process consist of combining all the solidity files, dependencies , imported libraries, and merging them into one continuous file. This include parsing the solidity files, resolve imports and dependencies to make sure that all the code in included in the flattened file.
    In the flattening process, we identify and include the functions within the main contract and the call relationships between them, including their interactions with interdependent functions from imported contracts. Comments are retained because they offer valuable context for understanding the intended functionality of the code. Moreover, the process ignores unreachable code for simplicity. Flattening a contract creates a single file that includes all code from the contract and its dependencies, eliminating the need to manage multiple files and ensuring that all dependencies are included correctly. To achieve this code flattening process, we utilized the Slither contract flattening tool \cite{feist2019slither}, specifically employing the OneFile strategy. Slither is a static analysis tool that detects potential issues in Solidity code, including problems related to imports and interfaces. The OneFile strategy merges all contracts within the project into a single, standalone file. By using Slither's OneFile strategy, we streamline the codebase and create a consolidated, easily accessible code structure for efficient analysis and development.

    \subsubsection{Function Call Graph (FCG) Extraction}
    \begin{algorithm}[t]
    \footnotesize
    
    \end{algorithm}

\begin{figure}
     \centering
     \includegraphics[width=\columnwidth]{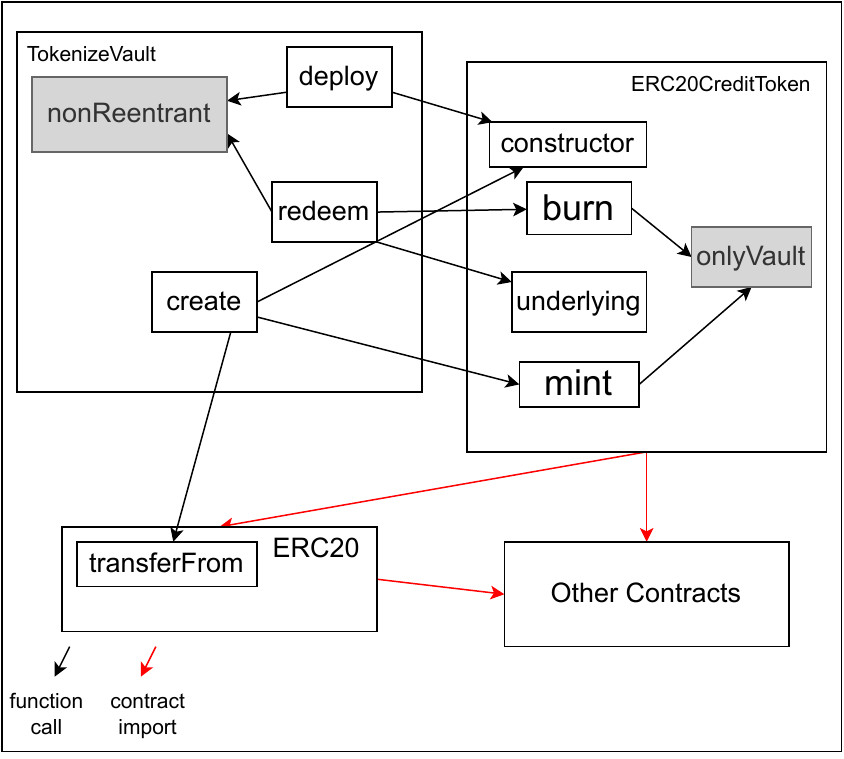}
     \caption{Function call graph of the tokenizeVault contract}
     \label{fig.2}
     \end{figure}

    The single file resulting from the flattening process is long and cannot fit within one prompt. One way to provide GPT with enough context is to focus on the Function Call Graph (FCG), which shows the flow of function calls, indicating which functions are called by other functions and how they are connected. FCG is utilized in our co-auditing approach to aid the segmentation mechanism to send only logically related code segments based on inter-dependencies rather than their physical location within the codebase. In our work, to generate a FCG from the flattened solidity file, we again used Slither to extract function calls from each contract. Slither uses the context of function calls and applies static and data flow analysis techniques to distinguish between internal and external function calls. The extracted function calls are then consolidated into a call graph.
    Based on the smart contract code in Section ~\ref{ME}, the generated FCG is shown in figure ~\ref{fig.2}.

    \subsubsection{CCL Generation}
    
    \begin{algorithm}
    \caption{CCL generation}
    \textbf{Input:} flattenedContract, adjacencyList \\
    \textbf{Output:} codeCallList
    %%\textbf{Input:} Flattened_contract , functions, edges\\
    %%\textbf{Output:} \textbf{code call list }
    \begin{algorithmic}[1]
    \STATE Initialize codeCallList as an empty dictionary
    \FOR{\textbf{each} functionId, callRelationships \textbf{in} adjacencyList}
        
        \FOR{\textbf{each} callRelationship \textbf{in} callRelationships}
        \STATE callerFunctionId, calledFunctionId  $\leftarrow$ callRelationship
        \STATE functionCode $\leftarrow$ \textbf{extractCode}(calledFunction, flattenedContract)
        \STATE codeCallList[functionId].append(functionCode)
        \ENDFOR    
    
    \ENDFOR
    \STATE \textbf{return}  codeCallList
    \end{algorithmic}
    \end{algorithm}

    Based on the FCG, we generate the CCL for each function. A CCL is a list containing all the function codes that are called during a specific function execution, which provides the GPT-4 code interpreter with a self-containing list of function codes. The number of CCLs corresponds to the number of functions implemented within the smart contract. Each CCL is composed of the code of the audited function, the caller function, and the code of the callee functions. This process is also recursive for the callee functions. This process guarantees that all the code being executed during a specific function call is included within the prompt. Our approach differentiates between external and internal function calls in the contracts. Using the function name and the type of the call, we locate and extract the exact called function and its corresponding code. This aims to provide the LLM with enough context to fully process the functions and recognize the vulnerability patterns.
    The listing \ref{lst2} illustrates a sample of CCL generated from the FCG for the function \textit{deploy}. 
    \begin{lstlisting}[caption= CCL list of the function 'deploy' extracted from contract TokenizingVault,
      label=lst2]
    ['function deploy( ERC20 underlying ) external nonReentrant returns (ERC20CreditToken) { // Create credit token if one doesnt already exist ERC20CreditToken creditToken = creditTokens[underlying]; if (creditToken == ERC20CreditToken(address(0))) { bytes memory tokenData = abi.encodePacked( underlying, address(this) ); creditToken = ERC20CreditToken( address(creditTokenImpl).clone(tokenData) ); creditTokens[underlying] = creditToken; } return creditToken; }', 
    'constructor(string memory name, string memory symbol, ERC20 _underlying) ERC20(name, symbol) { underlying = _underlying; vault = msg.sender; }',
    'modifier nonReentrant() { _guardCounter += 1; uint256 localCounter = _guardCounter; _; require(localCounter == _guardCounter); }']
    \end{lstlisting}
    
    \subsection{Reporting Scoping - CAQ Prompt Design}
    \label{sec:Prompts Phrasing Approaches}
    
    The effectiveness of LLMs in audit reporting heavily relies on how specifically the questions are posed. Accurate and relevant responses from LLMs depend on question precision, which is crucial in audit reporting, where delivering exact content in a certain order and format is essential. By formulating targeted questions, we direct the LLM to produce the required content accurately and focus on generating subsequent relevant information. Therefore, integrating a response template into the prompt is a key strategy to reduce response variability and standardize the output format of LLMs in audit reporting. In order to do this, we adopted a focus group methodology to identify the essential components of audit reports \footnote{\url{https://code4rena.com/reports/2024-03-dittoeth}}. Initially, a diverse collection of existing audit reports from our partner and from Code4rena\footnote{https://code4rena.com/reports} is assembled for analysis to understand the structure and content of audit reports. Then, a focused group of two professional auditors, in addition to three students, examined the reports. Each member of the group was asked to extract a schema of the main report parts. The group then agreed on a common schema that represents key report components that need to be included in the audit report. Based on the discussion, the group agreed that a report needs to specify  (i) if a vulnerability exists, (ii) a proof of the concept of the vulnerability, (iii) the business impact of the vulnerability, and (iv) suggestions for potential fixes. Building on this, we create the CAQ-based prompt strategy, then refine it to address Assessment Scoping, as we explain in the next section. The CAQ-based prompt strategy aims at eliciting specific, relevant responses in alignment with these identified components. Pilot testing and iterative refinement based on feedback from both the professional auditors and the students were conducted on a small dataset, leading to a finalized template that generated required, human-like responses. To limit the impact of the non-deterministic nature of generative models, we fixed the temperature in all tests to 0 in the GPT-code-interpreter during the pilot testing.
    Figure ~\ref{fig:CAQ_prompt} illustrates the CAQ-based prompt structure. To get a complete audit report based on this approach, the number of prompt executions is equal to the number of functions in a smart contract. 
    
    \begin{figure}[ht]
    \centering
    \begin{tcolorbox}
    \small
    Analyze the following code, respond me in this format  (fill in the part):
    
    1. Are there any  vulnerabilities? Provide a brief initial response with one of the following: ‘Yes,’ ‘No,’ or ‘Not sure.’ If your answer is ‘No,’ you can stop here. If your answer is ‘Yes,’ proceed to the next steps.
    
    2. Explain in details how this vulnerability can be exploited :\_ \_ \_. 
    
    3. The business impact of this vulnerabilities in one sentence: function leads to. 
    
    4.The potential solutions of this vulnerabilities. 
    
    Note: You can state more than one lines which have this vulnerability! This is a short extraction of the solidity code and please assume all the variables in this code have been defined.  
    \end{tcolorbox}
    \caption{CAQ-based prompt}
    \label{fig:CAQ_prompt}
    \end{figure}
    
    \subsection{Assessment Scoping - CWE Prompt Design}
    
    To incorporate context related to the expected target responses, we refined the reporting template by incorporating information about the common expected vulnerabilities and weaknesses.  To do so, we followed a similar approach to the work in \cite{david2023you}. In this work, the authors analyzed a set of 52 attacked Decentralized Finance protocols that have led to  952M USD in loss. The exploited vulnerabilities in the smart contracts have been categorized into 38 categories. In this paper, we use this list of the 38 most common vulnerability types to refine the prompt design. The aim is to make it vulnerability-specific. The LLM is prompted sequentially to identify only one type of vulnerability at a time from the CWE list. In the prompt, we request the LLM to respond with yes, no, or not sure to a \textbf{specific type} of vulnerability. This aims to enhance the guide for the attention mechanism of LLM. 
    
    Figure~\ref{fig:CWE_prompt} illustrates the structure for CWE-based prompts. For each function, we execute this prompt for all 38 vulnerabilities identified in the CWE. In order to generate a complete audit report for a smart contract, we repeat the execution of the prompt for every function in the contract. The total prompt execution required equals 38 multiplied by the number of functions in the contract. 
    
    \begin{figure}[ht]
    \centering
    \begin{tcolorbox}
    \small
    Analyse the following code, respond to me in this format  (fill in the part with \_ \_ \_): 
    
    1.Are there any \textbf{[CWE type]} vulnerabilities? Provide a brief initial response with one of the following: ‘Yes,’ ‘No,’ or ‘Not sure.’ If your answer is ‘No,’ you can stop here. If your answer is ‘Yes,’ proceed to the next steps.
    2. Explain in details how this vulnerability can be exploited :\_ \_ \_. 
    
    3. The business impact of these vulnerabilities in one sentence: \_ \_ \_ 
    
    4. The potential solutions of this vulnerabilities:\_ \_ \_
    
    Note: You can state more than one lines which have this vulnerability! This is a short extraction of the solidity code and please assume all the variables in this code have been defined.
    
    \end{tcolorbox}
    \caption{CWE-based prompt}
    \label{fig:CWE_prompt}
    \end{figure}
    
    \section{Evaluation}
    \begin{figure*}[t!]
    \centering
    \includegraphics[width=.9\textwidth]{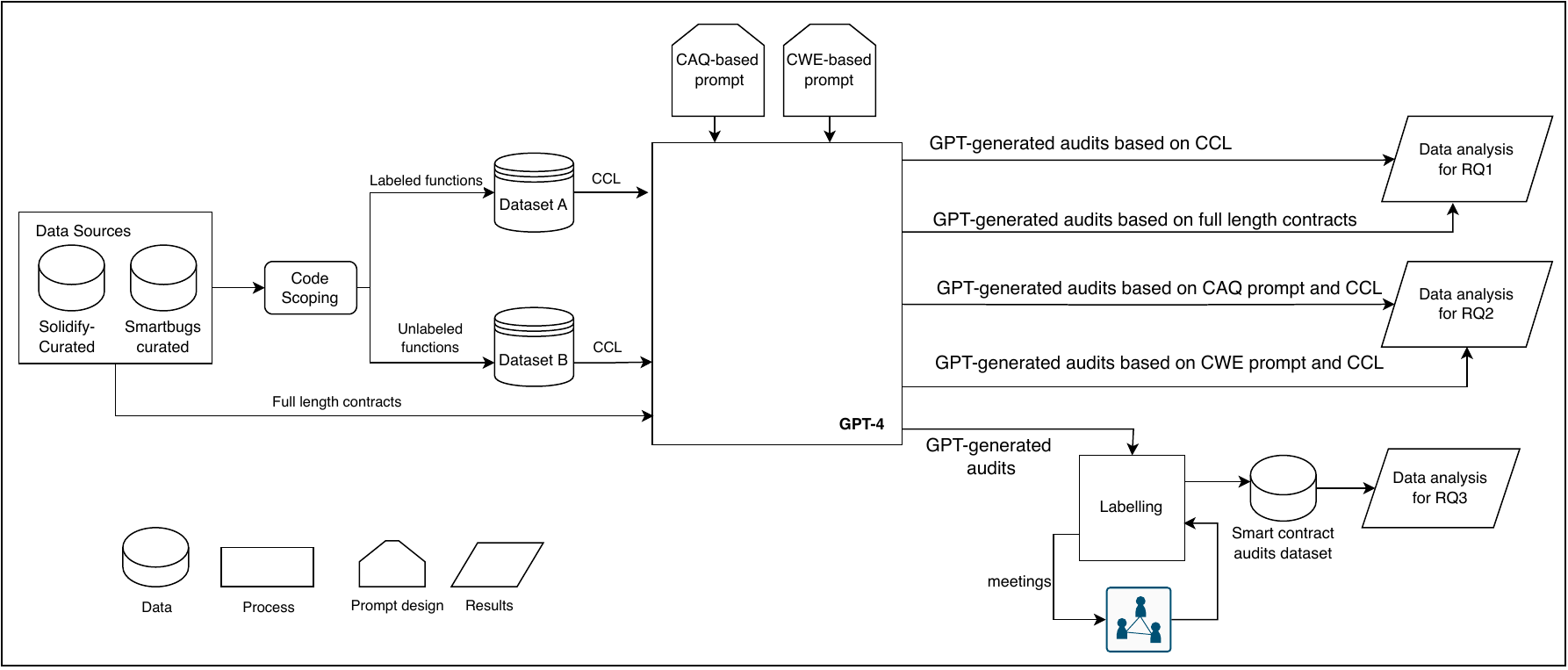}
    \caption{Overview of the empirical study approach}
    \label{ESD}
    \end{figure*}
    This section evaluates our proposed methods using an empirical study, detailed in Figure \ref{ESD}, which outlines our approach to addressing the research questions.
    
    \subsection{Data Sources}
    
    Our study utilizes two labeled datasets: Smart Bugs Curated \cite{durieux2020empirical} and SolidiFI-Benchmark \cite{ghaleb2020effective}. Smart Bugs Curated is an academic collection of 69 vulnerable smart contracts that are categorized into ten categories based on vulnerability types, such as access control, arithmetic, and bad randomness \cite{durieux2020empirical}. Each contract is annotated with the precise location and classification of its vulnerabilities. The SolidiFI-benchmark repository contains 50 vulnerable contracts covering seven bug types, including reentrancy, timestamp dependency, unhandled exceptions, and unchecked send. Each folder represents a specific vulnerability type. The repository also includes injection logs that identify the exact locations and types of bugs introduced in the code.
    
    The vulnerable smart contracts in both datasets are classified into different categories. Each smart contract belongs to only one category, and only one function of the smart contract is vulnerable. During our preliminary study, we noticed that GPT reveals different vulnerabilities in one smart contract. These vulnerabilities affect functions that are not claimed to be vulnerable.
    Moreover, although these datasets contain the labeled vulnerabilities, they might also include other vulnerable functions that are not labeled. To ensure the integrity of the study and given that GPT could have seen the publically available labeled datasets, we considered unlabeled functions in datasets. Hence, in this study, we divide the data sources into two datasets: \textbf{Dataset A} and \textbf{Dataset B}. Dataset A includes the labeled vulnerable functions, while Dataset B has unlabeled functions extracted from the data sources. 
    
    \subsection{Experiment Setup}
    \begin{table}[ht]
    \centering
    \caption{Experiment Sources}
    \label{tab:experiment_sources}
    \begin{tabular}{|c|c|c|c|}
    \hline
     Sources & Data & Prompt & source code \\ \hline
     source 1 & Dataset A & CAQ prompt & full code \\ \hline
     source 2 &Dataset A & CWE prompt & full code \\ \hline
     source 3 & Dataset A & CAQ prompt & CCL \\ \hline
     source 4 & Dataset A & CWE prompt & CCL \\ \hline
     source 5 & Dataset B &  CAQ prompt & CCL \\ \hline
     source 6 & Dataset B &  CAQ prompt & full code \\ \hline
    \end{tabular}
    \end{table}
    As mentioned in section~\ref{sec:methodology}, our approach is based on three scoping strategies: \textit{code scoping} aims to shorten the prompted code, \textit{reporting scoping}, provides GPT with a standardized output format for the audits, and \textit{assessment scoping}, sets a focus at the targeted vulnerability that GPT should address. These strategies were used to guide the design of different prompting templates. In our empirical study, we study the impact of employing these strategies on the GPT performance in smart contract code auditing.
    
    We designed six experiments to answer the research questions based on the different prompting templates and whether they are used with the full code or the Code Call List (CLL). Table~\ref{tab:experiment_sources} shows all the experiments; columns show the source name as it appears in the replication repository, the dataset used, the prompt template used, and whether the experiment is based on the full code of CCL, respectively.
    These experiments were used to answer the questions listed in section \ref{sec:motivation}
    \begin{table*}[!htb]
        \centering
        \caption{Dataset A experimental results}
        \label{tab:experiment_results}
        \begin{tabularx}{\textwidth}{l*{5}{>{\centering\arraybackslash}X}}
            \toprule
            Vulnerability type & \multicolumn{2}{c}{CAQ prompt} & \multicolumn{2}{c}{CWE prompt} & ground truth\\
            \cmidrule(lr){2-3} \cmidrule(lr){4-5} 
             & full contracts & CCL & full contracts & CCL  \\
            \midrule
            access control     & 11 & 21 & 17 & 21 & 23  \\
            bad randomness  & 9 & 26 & 10 & 29 & 30 \\
            denial of service   & 3 & 6 & 6 & 6 & 6\\
            reentrancy     & 23  & 32 & 23 & 32 &32\\
            arithmetic & 12  & 23 & 18 & 23 & 23 \\
            front running  & 3  & 6 & 4 & 6 & 6 \\
            unchecked low level calls  & 41  & 49& 41 & 50 & 54\\
            time manipulation  & 4  & 4 & 3 & 4 & 4\\
            short addresses  & 1 & 1 & 1 & 1 & 1\\
            other  & 2  & 5 & 2& 5 & 5\\
            \midrule
            total & 109 & 173 & 125 & 177 & 184 \\
            \bottomrule
        \end{tabularx}
    \end{table*}
    In this study, we utilized GPT-4's code-interpreter plugin with a fixed temperature of 0 and a token limit of 8000. Due to the absence of a GPT-4 API code-interpreter for code interpretation at that time, we employed Selenium—an open-source framework for automating web applications—to interact with the GPT-4 playground. This setup automated the process of generating and submitting prompts, as well as collecting responses. All the scripts we used for the experiments can be found in this repository this.\footnote{\url{https://zenodo.org/records/10530868}}
    
    \subsection{Empirical Study Results}

    \subsubsection{\textbf{Impact of the code scoping on the GPT performance}}\hfill
    \begin{tcolorbox}
    \textbf{RQ1.} How does CCL-based chunking perform in comparison to full code prompting?
    \end{tcolorbox}
    In this research question, we evaluate the performance of the proposed CCL chunking approach and compare it to the full code prompting performance using different prompt designs.
    Consequently, we have divided RQ1 into two more specific questions:
    \begin{itemize}
        \item RQ1.1: When utilizing the CAQ prompt, how does the performance of CCL-based chunking compare to that of full code prompting?
        \item RQ1.2: When utilizing the CWE prompt, how does the performance of CCL-based chunking compare to that of full code prompting?
    \end{itemize}
    
    After prompting GPT-4 interpreter using the different strategies, the responses are collected in CSV files and manually compared to the ground truth. A response is considered correct if GPT correctly identifies the same type and exact location of the vulnerability as described in the ground truth contracts. For each type of vulnerability, we report the number of correctly detected vulnerabilities as shown in table \ref{tab:experiment_results}. 
    
    To answer RQ1.1, we conduct hypothesis testing\cite{wohlin2012experimentation} on results from sources 1 and 3 in Table ~\ref{ESD}. We formulated the null hypothesis (H0) and the alternative hypothesis (H1) as follows:
    
    \noindent\textbf{RQ1.1-H0:} \textit{When utilizing the CAQ prompt, there is no statistically significant difference in the performance of CCL-based chunking compared to that of full code prompting.}
    
    \noindent\textbf{RQ1.1-H1:} \textit{When utilizing the CAQ prompt, there are statistically significant differences in the performance of CCL-based chunking compared to that of full code prompting.}

    To compare the efficiency of these strategies (i.e., CAQ prompt with CCL-based chunking vs. CAQ prompt with full code), which operate as classifiers, we use McNemar's test, which is particularly useful for comparing the performance of two classifiers. The test statistic measurement indicates a value of 62.01 following a chi-squared distribution with one degree of freedom.
    For a significance level of 0.05, consulting the chi-squared distribution table for one degree of freedom, we identified a critical value of 3.841. Hence, the test statistic is greater than the critical value; we can conclude that there is a significant difference in terms of effectiveness between the full code and CCL-based chunking methods.
    
    The detection rates are 59.24\% and 94.02\% for full code and CCL-based chunking methods, respectively. The Cohen's d value is equal to -0.45, which suggests that the difference in effectiveness between the two approaches is moderately significant. The stark contrast in these central tendencies, coupled with a notably low p-value of approximately 0.000287, strongly supports the rejection of RQ1.1-H0. Therefore, we conclude that when utilizing the CAQ prompt, there are differences in the performance of CCL-based chunking compared to that of full code prompting, with CCL-based chunking performing much better.
    
    To answer RQ1.2, we conducted a hypothesis test on the results from sources 2 and 4 as presented in Table ~\ref{tab:experiment_sources}. We formulated the null hypothesis (H0) and the alternative hypothesis (H1) as follows:
    
    \noindent\textbf{RQ1.2-H0:} \textit{When utilizing the CWE prompt, there is no statistically significant difference in the performance of CCL-based chunking compared to that of full code prompting.}
    
    \noindent\textbf{RQ1.2-H1:} \textit{When utilizing the CWE prompt, there are statistically significant differences in the performance of CCL-based chunking compared to that of full code prompting.}
    
    Similar to RQ1.1, we compare the efficacy of these strategies (i.e., CWE prompt with CCL-based chunking vs. CWE prompt with full code) by using McNemar's test. Applying the test statistic formula of McNemar's test indicates a value of 50.01, which is superior to the critical value obtained from the chi-squared distribution table with one degree of freedom, which is equal to 3.841. This value indicates a significant performance difference between the two approaches.
    The Cohen's d value is measured at -0.36, suggesting that the mean of correctly detected vulnerabilities is lower for full code prompting compared to CCL-based chunking. The substantial disparity between these means is robustly supported by a p-value of approximately 0.00117, surpassing conventional thresholds for statistical significance and compellingly justifying the rejection of the null hypothesis. The accuracy rates are 67.93\% and 96.02\%, respectively. Additionally, the sensitivity, F1-score, and precision are 76.69\%, 80.91\%, 85.62\% and 98.88\%, 98.06\%, 97.25\%, respectively. Consequently, we assert that under the CWE prompt, differences exist in the performance of CCL-based chunking and full code prompting, with CCL-based chunking demonstrating superior performance.
    
    Combining the conclusion from RQ1.1 and RQ1.2, we found that:
    \begin{tcolorbox}
    \textbf{Finding 1:} Under both prompt methods, the CCL-based chunking method performs better than the full code prompting.
    \end{tcolorbox}
    \subsubsection{\textbf{Impact of the assessment scoping on GPT performance}}\hfill
    
    \begin{tcolorbox}
    \textbf{RQ2.} What is the impact of different prompt phrasing
    approaches (CAQ-based or CWE-based) on the effectiveness of the GPT results of CCL-based chunking?
    \end{tcolorbox}
    
    Building upon the findings from RQ1, we focus on evaluating the CAQ-based and CWE-based assessment scoping methods, specifically when applied in conjunction with the CCL-based chunking approach. This experiment is designed to test the following hypotheses:
    
    \noindent\textbf{RQ2-H0:} \textit{Using CCL-based chunking, there is no difference in the proportion of found vulnerabilities between the CAQ prompt and the CWE prompt.}
    
    \noindent\textbf{RQ2-H1:} \textit{Using CCL-based chunking, there are differences in the proportion of found vulnerabilities between the CAQ prompt and the CWE prompt.}
    
    In order to assess and compare the performance of the CAQ and CWE prompts, we calculate McNemar’s test \cite{pembury2020effective}, which results in a value equal to 2.25. This value is inferior to the critical value obtained from the chi-squared distribution table for 1 degree of freedom and a 0.05 confidence significance level. Using this metric, we are not able to reject the null hypothesis, RQ2-H0. The statistical analysis yielded a p-value of approximately 0.553, which fails to provide substantial evidence to reject the null hypothesis, indicating no significant difference.
    The accuracy rates are 94.02\% and 96.02\%, respectively. the recall value is equal to 94.02\% and 96.02\%.
    To calculate the effect size, we first calculate the standard deviation for the CAQ and CWE prompts, which results in 15.61 and 16.04, respectively. The Cohen's d value resulted in -0.025, which is considered very small. This value indicates that there is a very small difference between the two approaches in terms of their ability to detect vulnerabilities. Additionally, the sensitivity, F1-score, and precision are 76.69\%, 80.91\%, 85.62\% and 98.88\%, 98.06\%, 97.25\%,
    
    Consequently, we conclude that, when using CCL-based chunking, there is no statistically significant difference in the proportion of discovered vulnerabilities between the CAQ and CWE prompts.
    We followed a specific methodology for our assessment as detailed in Section III.C. For each function, we utilized the CAQ prompt once, while the CWE prompt was executed for all 38 vulnerabilities identified in the Common Weakness Enumeration (CWE). Notably, any vulnerabilities not defined in CWE would not be detected by the CWE prompt, potentially leading to false negatives. Therefore, we assert that:
    
    \begin{tcolorbox}
    \textbf{Finding 2:} In the context of comparing the performance of the CAQ-based prompt and the CWE-based prompt, it is observed that there is a minimal disparity in their efficacy. However, it is worth noting that the CAQ-based prompt demands less effort and can detect a wider range of vulnerability types. 
    \end{tcolorbox}
    
    Integrating the insights derived from RQ1 and RQ2, we can conclude that:
    
    \begin{tcolorbox}
    \textbf{Finding 3:} The most effective approach for detecting vulnerabilities in Solidity code using the GPT-4 code interpreter, within the context of both code segment and text description prompts, is the combination of CCL-based chunking and the CAQ-based prompt.
    \end{tcolorbox}
    We have redone the tests on the redeem function that we explained in the section 2.3, using the CAQ-based prompt along with CCL, GPT4 was able to detect effectively the vulnerability  
     \footnote{https://chat.openai.com/share/fb51d632-c6aa-47b0-a81e-7ec473c28b92}.
    \subsubsection{Human qualitative evaluations}\hfill
    \begin{tcolorbox}
    \textbf{RQ3.}   To what extent does a human auditor analysis of GPT-4 code interpreter’s results on an unlabeled dataset align with our previous findings?
    \end{tcolorbox}
    Addressing RQ3, we aim to evaluate the reliability and add an extra layer of validity to the proposed approach. We conducted a qualitative experiment in which we asked expert human auditors to audit and manually validate the auditing reports generated by our approach on a set of unlabeled smart contract functions. This evaluation was conducted under the assumption of a low likelihood that GPT models were previously exposed to, or trained on, the specific function annotations (labels). We hope this will provide insights into how well our approach performs in scenarios likely unfamiliar to the GPT's training data. The following sections detail the dataset, the experiment, and the insights from validation.

    \textbf{Data Collection:}
    For this experiment, we used Dataset B, which includes the CCL of 86 functions. These functions were selected from real-world smart contracts. Our selection process involved choosing complex (long) and diverse smart contracts, from which we extracted functions not previously labeled as vulnerable. This selection strategy was based on the presumption that the labels (vulnerabilities) of these unlabeled functions were less likely to be part of GPT's training set, ensuring an unbiased and rigorous assessment. The initial results (Finding 1) indicated that CCL-based chunking effectively reduced the size of the prompted source code. However, this leads to an important question: does the public availability of datasets enhance GPT's ability to detect vulnerabilities?

    Dataset B comprises unlabeled functions paired with corresponding Contract Contextual Labels (CCL). To ensure rigor, we selected contracts meeting specific criteria:
    \begin{itemize}
        \item Inter-contract dependencies (importing libraries/contracts): 
        This criterion ensures the dataset captures scenarios where functions interact with external libraries or contracts, reflecting real-world complexity.
        \item Complex logic and lengthy code (more then 100 lines):
        This criteria is used to ensure complex contracts functionalities.
        \item Varied contract lengths (representative sampling):
        Including contracts of different sizes ensures the dataset covers diverse use cases, providing a balanced evaluation of model performance across both simple and complex scenarios.
        \item Developer-commented functions (simulating human auditing):
        Human auditors generally do audits to commented contracts. The comments help auditor to understand the code functionality and thus help him detect logic vulnerabilities.
    \end{itemize}
    This diversified dataset challenges Large Language Models (LLMs) and mirrors real-world auditing scenarios.

    \textbf{Annotation and Evaluation}
    
    To qualitatively evaluate the results, the auditors manually annotated the GPT-4-code-interpreter's responses using two criteria; namely, response correctness and relevance:
    
    \textit{Response Correctness} measures whether the judgment of GPT is correct regarding the existence of the vulnerability and its location on the code. We noticed that when GPT identifies the type of vulnerability correctly, it also locates the vulnerable location correctly. Correctness is divided into three sub-categories:
    \begin{itemize}
        \item {level 1 (NOT) :} GPT response is not correct
        
         \item {level 2 (Partial) :} The generated audits successfully detect at least one correct vulnerability.
        
         \item{level 3 (Perfect) :} All the vulnerabilities included in the GPT audit are correct.
    \end{itemize}
    
    \textit{Response Relevance} measures the quality of explaining how the vulnerability can be exploited. It reflects the clarity of the response and how comparable it is to the human-generated audits. The response relevance is divided into three categories:
    \begin{itemize}
        \item {level 1 (NOT) :} The response is not relevant compared to a manual audit.
        
         \item {level 2 (Partial) :} The response includes some relevant information that can be helpful to a professional auditor.
    
         \item {level 3 (Total) :} GPT response is similar to manual audits.
    
    \end{itemize}

    We employed a rigorous annotation process for the manual study of GPT-generated audits of the selected 86 CCLs. Two experienced auditors from our industrial partner annotated the samples independently along two dimensions. The annotators have extensive expertise in the development and auditing of smart contract projects. The process of annotating the GPT responses was conducted in rounds. First, each auditor independently reviewed and evaluated the GPT responses based on the correctness and relevance of the responses. Then, a meeting between the two annotators and the co-authors was held to ensure they shared the same understanding of the task. The goal was to enable the annotators to perform their tasks with a well-defined framework and to enhance the quality and consistency of their individual assessments, recognizing that differences in judgment are inherent in this challenging domain.
    The annotators were able to access the full source code corresponding to each CCL to have an accurate understanding of the functions being audited. Different meetings were held between the auditors to resolve discrepancies between annotations. The final Cohen's Kappa coefficient \cite{mchugh2012interrater} for response correctness and response relevance was 0.65 and 0.68, respectively, suggesting moderate agreement between the two annotators. Even though the relevance score between the two annotators is not exceptionally high, the labeling results retain their value. This is because our analysis focuses on comparing the changes in scores assigned by the same annotator to the CCL-based chunking and full code usage performances rather than making a direct comparison between the scores from two annotators. This approach underscores the individual annotator's consistency and personal benchmark standards, providing a more controlled and internally consistent comparison.
    
    \begin{figure}[!htb]
        \centering
        \subfloat[Response relevance]{\includegraphics[width=0.45\linewidth]{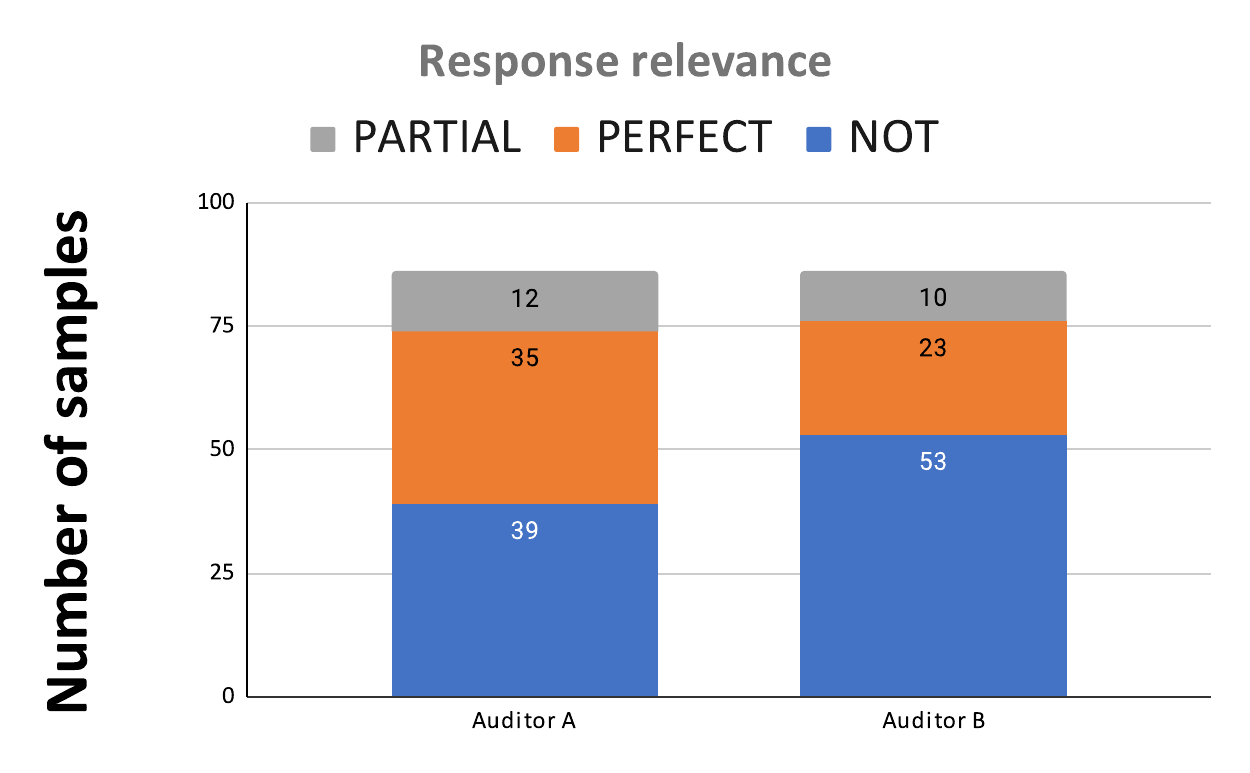}\label{fig:figure1}}
        \hfill
        \subfloat[Response correctness]{\includegraphics[width=0.45\linewidth]{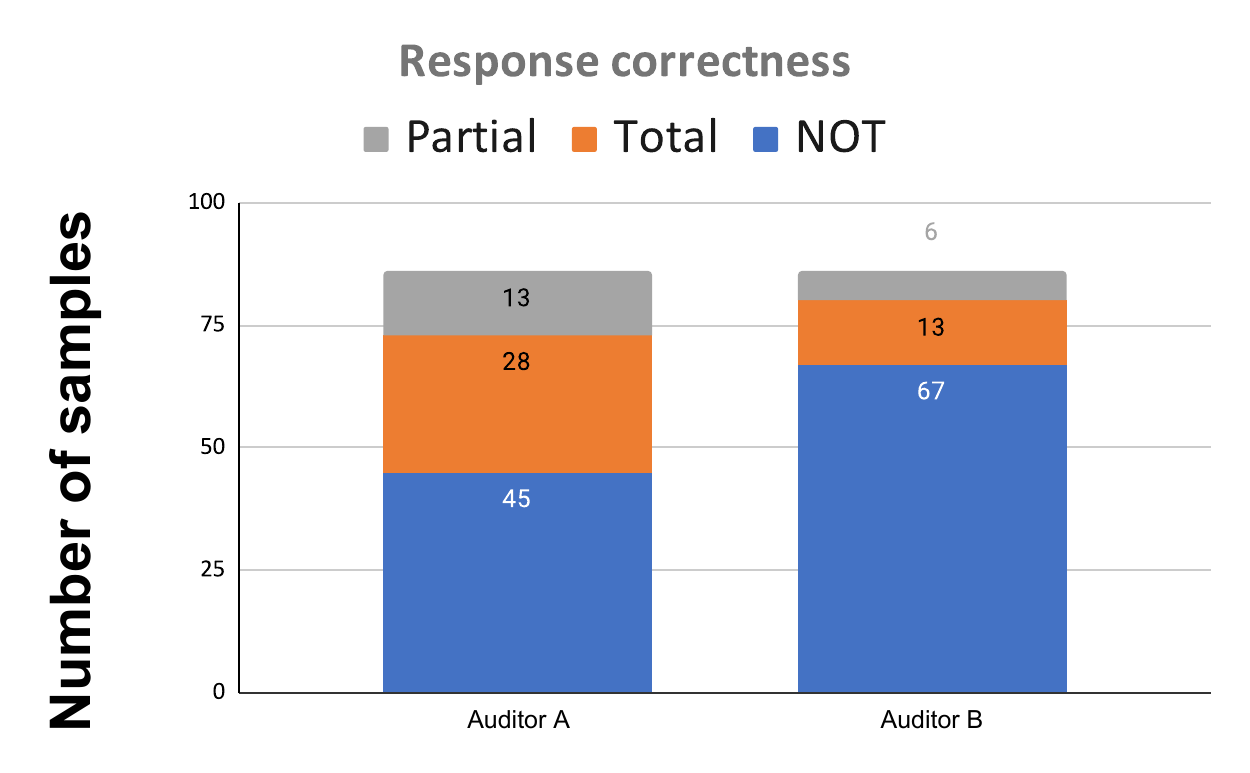}\label{fig:figure2}}
        \caption{Qualitative evaluation of the GPT generated audits of two reviewers}
        \label{fig:annotation_results}
    \end{figure}
    
    \textbf{Results:} The outcomes of the annotation procedure are depicted in Figure \ref{fig:annotation_results}. The results show that despite using unlabeled datasets, on average, expert auditors concurred that with our prompting approach, the GPT-4 interpreter correctly and accurately generated reports in 56\% of the cases, and these reports were relevant in 46\% of instances. These percentages increase to 76.5\% and 75\% for partial correctness and relevance, respectively. This aligns with our prior findings that the application of CCL-based chunking enhances the GPT-4 code interpreter's ability to identify vulnerabilities in unlabeled datasets. The derived auditing dataset can be accessed openly in the project's repository. \footnote{\url{https://zenodo.org/records/10530868}}

    \begin{tcolorbox}
    \textbf{Finding 4:} Analysis from human annotations suggests that, in unlabelled datasets, the use of CCL-based chunking improves the capability of the GPT-4 code interpreter in identifying vulnerabilities.
    \end{tcolorbox}

    \section{Discussions}
    In this section, we delve into a comprehensive discussion of the methodologies employed in our study and their implications. By reflecting on the strengths, challenges, and nuances of each method, we aim to provide insights that can inform future research and practical applications in the domain of code vulnerability assessments.
    
    \textbf{A. Affirming the Efficacy of Our CCL-Based Chunking Approach}
    
    Our CCL-based chunking method, rooted in the function call graph (FCG) framework, is indicative of  efficiency in interconnecting related functions at different positions in the code. This approach ensures that related functions are consolidated within the same chunked block. Despite the code being segmented, the logical association among related functions is preserved. This technique not only shortens the length of code fed into each prompt, thus addressing the token limitations but also maintains the code's logical continuity, crucial for understanding the broader context. We attribute the success of the CCL-based chunking, in terms of handling code length issues and achieving commendable accuracy, to these methodological strengths.
    
    To compare with the static analysis tools, we run multiple vulnerability detection tools on the smartBugs-curated dataset. The results are shown in table \ref{tab:vulnerability_detection_tools}. 
    The table compares the performance of the proposed CAQ+CCL (GPT + context augmenting) approach with well-known vulnerability detection tools: Slither\cite{feist2019slither}, Mythril\cite{sharma2022survey}, Oyente\footnote{https://github.com/enzymefinance/oyente}, and Surya\footnote{https://github.com/Consensys/surya}. These tools represent traditional auditing methods, with Slither and Oyente focusing on static analysis, Mythril combining static and dynamic analysis with symbolic execution, and Surya providing static analysis along with code visualization. The table shows that CAQ+CCL outperforms all these tools in terms of the number of vulnerabilities detected, with a detection rate of 83.57\%. In comparison, Slither detected 102 vulnerabilities with a rate of 49.27\%, Mythril identified 100 vulnerabilities with a rate of 48.30\%, Oyente detected 90 vulnerabilities with a rate of 43.47\%, and Surya only identified 21 vulnerabilities with a rate of 10.14\%. This highlights the superior performance of CAQ+CCL, not only in detecting more vulnerabilities but also in providing explanations of how these vulnerabilities can be exploited, offering more comprehensive support for auditors compared to the other tools.
    
    \begin{table}[ht]
    \centering
    \caption{Comparison with Vulnerability Detection Tools}
    \label{tab:vulnerability_detection_tools}
    \begin{tabularx}{\textwidth}{|X|X|X|X|X|}
    \hline
    \textbf{Tool Name} & \textbf{Type} & \textbf{Number of Detected Vulnerabilities} & \textbf{Detection Rate (\%)} & \textbf{Ground Truth} \\ \hline
    CAQ+CCL approach & GPT + context augmenting & 173 & 83.57 & 207 \\ \hline
    Slither & Static analysis & 102 & 49.27 & 207 \\ \hline
    Mythril & Static and Dynamic Analysis and symbolic execution & 100 & 48.30 & 207 \\ \hline
    Oyente & Static Analysis & 90 & 43.47 & 207 \\ \hline
    Surya & Static Analysis and Code Documentation & 21 & 10.14 & 207 \\ \hline
    \end{tabularx}
    \end{table}

    \textbf{B. Adequacy of Our CAQ-Prompt in Practical Application} \hfill
    
    Our meticulously designed CAQ prompt is structured around four core components: judgment of vulnerabilities, explaining how the vulnerability can be exploited, business impact, and potential solutions. This structure ensures that the responses obtained are well-organized and present the comprehensive information one would expect in a professional audit report, thereby laying the groundwork for LLMs to generate thorough audit reports in the future.
    
    Furthermore, by comparing the CAQ-based prompt with the traditional CWE-based prompt, we observed that the overall accuracy of the CAQ prompt is competitive, often requiring significantly less effort. For instance, in \cite{david2023you}, the authors identified 38 types of vulnerabilities, implying that using a CAQ prompt could reduce the workload by up to 39 times compared to employing CWE-specific prompts.
    Given these advantages in terms of both efficiency and breadth of detection, we advocate for adopting CAQ-based prompts in LLM-based code vulnerability assessments.

    \textbf{C. Interpretation of the effectiveness of the CCL based chunking technique }
    
    The proposed approach aims to provide an efficient prompting techniques that aims to reduce the prompt sizes specially regarding complex codebase while providing GPT with enough data in order to analyze effectively the targeted functions and localize vulnerabilities. As discussed by authors in \cite{levy2024same}, As the input length increases, the model has to manage a larger token numbers, which can lead to difficulties in maintaining coherence and relevance. This is particularly challenging in code-related tasks where precise understanding and execution of instructions are crucial. While models like GPT-4 have a high token limit, they still face challenges when approaching this limit. The need to chunk inputs or manage tokens effectively can result in loss of context or important details, impacting the quality of generated audits. 
    
    Besides, GPT4 model rely on the self-attention mechanism to understand the relationships between words and their contexts. Self-attention allows each token to "attend" to all other tokens in the input sequence. However, this attention mechanism scales quadratically with the sequence length (\( O(n^2) \) complexity). For long sequences, the amount of computational resources and memory required increases exponentially, making it harder for the model to effectively capture relationships \cite{wang2024beyond}.
    
    As discussed in the section 4.3.2, the finding 2 and 3 suggest that there is minimal  disparity between the two prompting technique and thus the most effective approach is the combination
    of CCL-based chunking and the CAQ-based prompt. This can be explained by pattern recognition employed by GPT4. GPT-4 is trained on vast datasets and is designed to recognize patterns and generalize information. Whether using CCL-based or CAQ-based prompting to find vulnerabilities in a smart contract, the llm applies its understanding of common vulnerabilities and patterns it has learned for the training data.
    
    The architecture of GPT models allows them to process and generate language based on learned patterns rather than specific rules. This means that for tasks like vulnerability detection, the model relies heavily on its training data's statistical properties rather than a deterministic set of rules for pattern recognition \cite{chen2023chatgpt}.

    \textbf{D. Interpretation of the performance of GPT using CWE and CAQ prompt }
    
    The observed similarity in performance between CAQ prompts (generic prompts focusing on all vulnerabilities) and CWE prompts (specific prompts targeting particular vulnerabilities) can be attributed to the fundamental architecture and operation of large language models (LLMs). Transformer-based LLMs, such as OpenAI Codex, rely on identifying and applying patterns learned during pre-training to the input context. Generic prompts likely activate a broad range of the model’s knowledge, encompassing both general and specific vulnerability patterns. This aligns with findings by \cite{chen2021evaluating}, which demonstrated Codex’s ability to handle diverse programming tasks, revealing that generic queries often leverage the model’s extensive knowledge base effectively.
    
    The retrieval mechanisms of LLMs further explain this overlap. Models tend to retrieve contextual patterns based on the input query while inherently including related peripheral information due to the interconnected structure of their training data. Even when tasked with identifying specific vulnerabilities, the model’s retrieval process draws upon a broader set of associated patterns. This phenomenon has been highlighted in studies like \cite{yuan2024towards}, which explore the dense interconnections within LLM training data, facilitating the retrieval of comprehensive knowledge irrespective of prompt specificity.
    
    Additionally, prompt optimization research sheds light on this behavior. Studies such as \cite{cheng2023black} indicate that carefully crafted prompts can enhance performance in tasks requiring deep reasoning or nuanced outputs. However, tasks that depend heavily on pre-trained knowledge, such as vulnerability detection, tend to exhibit minimal sensitivity to prompt specificity. Generic prompts already activate the broad retrieval patterns necessary to address most specialized queries, resulting in limited differences in performance between general and specific prompts.
    
    These findings suggest that LLMs exhibit a degree of prompt invariance, particularly in domains where their pre-trained knowledge is expansive and well-integrated. For practical applications like smart contract auditing, this insight implies that focusing on robust generic prompts may suffice, reducing the need for extensive specialized prompt engineering. By leveraging the inherent strengths of LLMs, such an approach can streamline workflows while maintaining high task performance.

    \textbf{E. Implications on the proposed approach on industrial use}
    
    The primary objective of the proposed methodology is to facilitate expert security audits of smart contract projects. By leveraging CAQ-based prompts and CCL-based chunking, auditors receive a comprehensive assessment of potentially vulnerable functions, including explanations of proof-of-concept exploits and corresponding mitigation strategies. This structured information enhances the auditor's understanding of individual function purposes within the broader contract and illuminates various exploitation pathways.
    
    Experimental results demonstrate that CCL-based chunking effectively reduces the volume of code required for LLM prompts while maintaining sufficient context for accurate interpretation. This technique scales effectively to support professional auditors working with large code repositories.

    \section{Related Works}
    
    Several approaches have been developed to detect vulnerabilities in solidity smart contracts using both static and dynamic analysis.
    
    Static analysis involves examining the contract's code without executing it. The goal is to identify vulnerabilities by analyzing the code's structure, syntax, and logic. Tools such as Mythril, Manticore\cite{mossberg2019manticore}, and Slither\cite{feist2019slither} are widely used for static analysis of Solidity code. Mythril, for example, uses static analysis, symbolic execution, and fuzzing for the detection of vulnerabilities. It also performs control flow and data flow analysis  to understand how data moves through the contract. Its static analysis approach involves disassembling ethereum smart contract bytecode to an intermediate representation, typically leveraging an abstract syntax tree (AST). This analysis is augmented with pattern matching and heuristic techniques to identify known vulnerability patterns such as reentrancy, integer overflow and underflow, and other smart contract vulnerabilities. nevertheless, these tools produce a high ratio of false positive/negative .
    
    Dynamic analysis, on the other hand, entails observing the contract's behavior during its execution within a controlled environment. The objective is to unearth vulnerabilities that might not be evident in a static examination but manifest during run-time. Instruments like Ganache and Truffle Suite are frequently employed for this purpose in the Solidity ecosystem. By simulating transactions and monitoring contract interactions in real-time, dynamic analysis can spotlight issues like gas inefficiencies, unpredictable state changes, and faulty event emissions, which might evade detection during static scrutiny. One of the challenges of using dynamic analysis is that it can be time-consuming and resource-intensive
    
    Another approach involves using formal verification and symbolic execution techniques \cite{dwivedi2019formal, wang2024efficiently} to mathematically prove the correctness of the smart contract. Formal verification involves using mathematical models to check the correctness of the code. This approach can be used to identify logical errors and ensure that the smart contract behaves as intended\cite{almakhour2020verification}. 
    
    Recently, the application of machine learning and deep learning techniques to analyze Solidity smart contracts has gained significant traction \cite{mi2021vscl, qian2022multi}. Deep learning models, especially Recurrent Neural Networks (RNNs) and their variant, Long Short-Term Memory (LSTM) networks, have been employed to capture the sequential nature of code and its inherent patterns \cite{qian2022multi}. LSTMs, with their ability to remember long-term dependencies, are particularly suited to recognizing intricate patterns in Solidity code that might be indicative of vulnerabilities. While this approach has shown outstanding results, its generalization to unseen smart contracts written in a new version of Solidity remains a concern. Also, these models are not able to generate human-like explanations of the found vulnerabilities and the mitigation steps.
    On the other hand, GPT has attracted a lot of attention from the research community regarding its ability to code review and generation \cite{sun2024gptscan, boi2024vulnhunt, zeng2023solgpt, ma2024combining, napoli2023evaluating}. Particularly in \cite{chen2023chatgpt}, a comprehensive evaluation of GPT-3.5 and GPT-4 code interpreter for vulnerability detection in smart contracts is illustrated. This work suggests using prompt engineering for the auditing of smart contracts. In particular, it first uses GPT to generate the prompt that describes the task of auditing; then, this prompt is used along with the full source code of the contracts as input to GPT. To measure the performance of GPT, the work considered only the Precision, Recall, and F-measure metrics. However, no qualitative evaluation of the generated output was made. The work did not consider the non-deterministic nature of the GPT model regarding the prompt design. Also, this work did not evaluate the capacity of GPT for the task of explaining the found vulnerabilities and the mitigation steps. The experiments showed that the prompt length, i.e., the contract length, can affect negatively the detection performance when exceeding the token limit.

    Similarly, in \cite{david2023you} a quantitative evaluation of the performance of GPT-4 code interpreter for the task of vulnerability detection has been performed. Particularly, the prompt design included a description of a specific attack type besides the whole source code of the contract. Common decentralized finance attacks have been classified into 38 Common Weakness Enumeration (CWE), and each smart contract has been prompted along with a description of the specific attack. Although the results of the experiment have shown great potential for GPT-4 code interpreter for detecting vulnerabilities, this experiment has shown that GPT performance is inversely proportional to contract length. 
    The authors in \cite{ma2024combining} suggest combining fine-tuning and LLM-based agents to detect and explain vulnerabilities in smart contracts. It employs a multi-step process where the model first identifies potential issues, then reasons about the underlying causes, and finally ranks and critiques these explanations to ensure the most accurate and justified results.
    
    In \cite{wang2024efficiently}, authors focused on using static analysis techniques along symbolic execution for efficient vulnerability detection in smart contracts. By analyzing the inter-contract program dependency graph and performing symbolic execution on the smart contracts, they focus on the detection of reentrency vulnerabilities. While the evaluation of this approach have shown its efficiency for reentrency vulnerability detection, the work has not evaluated the approach on other types of smart contracts specific vulnerabilities. The authors in \cite{wei2024llm} present an innovative framework that utilizes a multi-agent conversational approach for detecting and analyzing vulnerabilities in smart contracts. This approach incorporates two primary strategies: Broad Analysis and Targeted Analysis. Although the evaluation demonstrates enhanced effectiveness in identifying vulnerabilities, the tool is highly resource-intensive. 
    
    Additionally, its performance can degrade when analyzing longer code segments. In \cite{mothukuri2024llmsmartsec}, authors suggest an approach leveraging fine-tuned GPT-4 to identify and fix vulnerabilities in Solidity code by creating an annotated Control Flow Graph (CFG)
    
    \section{Threat to validity}
    
    \textbf{Internal Threat:} A potential threat to the internal validity of our study arises from our automated script designed for extracting the Code Call List (CCL). Particularly when importing interfaces, complications may emerge. Due to various possible implementations for the same function signature, our algorithm might erroneously select an incorrect imported function. To alleviate this concern, we meticulously cross-examined all constructed CCLs, ensuring the correct implementation for each function was chosen.
    In real-world applications, smart contracts are often commented. The comments serve as a documentation of the written code. It also helps to enhance the code readability and clarity, which provides the developers and auditors with a deeper understanding of the aimed functionality of the code. For all these reasons, we did not remove the comments from the code. Our evaluation is performed on commented code. However, different smart contracts, including those obtained through reverse engineering, may not have comments, which may decrease the performance of the proposed approach.
    Another aspect to consider is the type of variables. Since the CCL is composed of a list of functions which can belong or not to the same contract, a special attention should be given to the variables being used inside the functions. Functions belonging to different smart contracts can use the same name for state variables, and GPT is unable to distinguish the difference between the two variables since it does not have access to the state variable declaration area. In future works, we will be prompting that section along with the functions list.
    
    \textbf{External Threat:} Our evaluation leverages data from sources such as Smart Bugs and the SolidiFy-benchmark. This data encompasses labeled vulnerabilities and has been open-source since 2020. Given that this data might have been incorporated into GPT-4 code interpreter's training, it poses a potential external threat. To mitigate this, we created a new dataset composed of unlabeled functions from the evaluation contracts. However, a lingering uncertainty remains: these functions might have been labeled by other datasets, though we have no confirmation of such overlaps.
    Dataset B mitigates the risk of model performance bias due to training overlap and enhances the evaluation’s real-world relevance by providing unseen smart contract code for auditing. The unseen smart contracts functions have been labeled by the two professional auditors from Quantstamp.
    
    In our study, we define the divide the context into three components which are : code scoping, reporting scoping and assessment scoping.
    The context window is defined as the maximum number of tokens that an llm can process in a single prompt. As code length increases, the complexity of interactions between different parts of the code also increases. LLMs performance for the task of auditing smart contracts decrease especially for long code chunks.
    \cite{levy2024same}. For this work, as the length of prompt is very important, we define  the context of code as the code chunks and dependencies that are required for execution of the specific function we are auditing. While other works may have a different definition of the context, for this work we limit our definition of the context to the code chunks and dependencies a specific function requires to execute correctly
    The non-deterministic nature of Generative Pre-trained Models (GPTs) poses significant challenges and limitations, particularly in contexts where consistency and reliability are paramount. This inherent unpredictability arises from the models' training on diverse and extensive datasets, which leads to a stochastic approach in generating responses. While this can yield creative and varied outputs, it also results in a lack of consistency, making the model's reliability questionable in critical applications.
    Slither is used for code flattening and function call graph construction in the suggested approach. Slither supports one file flattening. We presume the created file and dependency management process are error-free. Complex dependencies might cause issues while flattening, resulting in unreachable code. All contracts that Slither fails to flatten are discarded.

    \section{Conclusion}
    
    In this paper, we introduce three context-augmentation approaches to improve smart contract co-auditing with Large Language Models (LLMs) and design prompts around these strategies. The proposed approaches address challenges in handling long code, augmenting audit process knowledge, and focusing on specific vulnerabilities or weaknesses, thus enhancing LLM effectiveness in smart contract co-auditing. Our empirical study shows that the CCL-based chunking method consistently outperforms full code prompting. In the realm of prompt design, a nuanced comparison between the CAQ-based and CWE-based prompts indicates that while both exhibit comparable proficiency, the CAQ-based prompt is more efficient and showcases a broader vulnerability detection scope. Notably, the most effective strategy for auditing vulnerabilities in Solidity code using the GPT-4 code interpreter emerges as a mix of CCL-based chunking with the CAQ-based prompt.

    \bibliographystyle{elsarticle-num}

    \bibliography{sample_base}
    \end{document}